\documentclass[10pt,journal]{IEEEtran}

\usepackage[cmex10]{amsmath}
\usepackage{array}
\usepackage[utf8x]{inputenc}
\usepackage{amsmath,amsfonts,amssymb,amscd,amsthm,xspace}
\usepackage{lipsum}
\usepackage{url}
\newcommand{\PR}[1]{\ensuremath{\left[#1\right]}}
\newcommand{\PC}[1]{\ensuremath{\left(#1\right)}}
\newcommand{\chav}[1]{\ensuremath{\left\{#1\right\}}}
\newcommand{\PM}[1]{\ensuremath{\left|#1\right|}}
\newcommand{\norm}[1]{\left\lVert #1 \right\rVert}
\usepackage{blindtext}
\usepackage{amsmath,amssymb,amsfonts}
\usepackage{graphicx}
\usepackage{textcomp}
\usepackage{xcolor}
\usepackage{algorithm}
\usepackage[noend]{algpseudocode}
\usepackage{lipsum}
\usepackage{comment}
\usepackage{makecell}
\usepackage{adjustbox}
\usepackage{multirow}
\usepackage{setspace}
\definecolor{r}{rgb}{0,0,0} 
\definecolor{red}{rgb}{0,0,0} 

\makeatletter
    \def\tagform@#1{\maketag@@@{\normalsize(#1)\@@italiccorr}}
\makeatother

\begin{document}
\bstctlcite{IEEEexample:BSTcontrol}
    \title{Robust MMSE Precoding and Power Allocation for Cell-Free Networks}
\author{Victoria M. T. Palhares, Andre R. Flores and Rodrigo C. de Lamare  \\
$^1$ ~Centre for Telecommunications Studies, Pontifical Catholic
University of Rio de Janeiro, Brazil \\
$^2$ ~Department of Electronic Engineering, University of York,
United Kingdom \\
Emails: victoriapalhares@cetuc.puc-rio.br, andre.flores@cetuc.puc-rio.br, delamare@cetuc.puc-rio.br.
\thanks{Copyright (c) 2015 IEEE. Personal use of this material is permitted. However, permission to use this material for any other purposes must be obtained from the IEEE by sending a request to pubs-permissions@ieee.org.}
}

{}


\maketitle


\begin{abstract}
We consider the downlink of a cell-free massive multiple-input
multiple-output (MIMO) system with \textcolor{red}{single}-antenna
access points (APs) and single-antenna users. An iterative robust
minimum mean-square error (RMMSE) precoder based on generalized
loading is developed to mitigate interference in the presence of
imperfect channel state information (CSI). An achievable rate
analysis is carried out and optimal and uniform power allocation
schemes are developed based on the signal-to-interference-plus-noise
ratio. \textcolor{r}{An analysis of }the computational
cost\textcolor{r}{s} of the proposed RMMSE and existing schemes is
\textcolor{r}{also} presented. Numerical results show the
improvement provided by the proposed RMMSE precoder against linear
minimum mean-square error, zero-forcing and conjugate beamforming
precoders in the presence of imperfect CSI.
\end{abstract}

\begin{IEEEkeywords}
Cell-free massive MIMO, robust MMSE precoding, power allocation, distributed antenna systems. \vspace{-0.5em}
\end{IEEEkeywords}

%
\IEEEpeerreviewmaketitle



\section{Introduction}
\label{introduction}

\IEEEPARstart{C}ell-free massive multiple-input multiple-output
(MIMO) systems have emerged in recent years as a combination of
massive MIMO, distributed antenna system\textcolor{r}{s} (DAS) and
network MIMO \cite{mmimo,wence}, where many randomly distributed
access points (APs) are connected to a central processing unit (CPU)
and serve simultaneously a much smaller number of users. At the CPU,
precoding and power allocation algorithms are performed. Cell-free
concepts have been shown to increase energy efficiency (EE) and
per-user throughput over cellular systems in rural and urban
scenarios \cite{Marzetta2016,Ngo2017,Nguyen2017}. Moreover, they can
have simple signal processing thanks to  favorable propagation with
channel hardening \cite{Marzetta2016}. Precoding is a key
interference mitigation technique
\cite{Spencer1,Stankovic,Sung,Zu_CL,Zu,wlbd,rsbd,rmmse,wlbf,locsme,okspme,lrcc,mbthp,rmbthp,rsthp,bbprec,baplnc,jpba,zfsec,rprec&pa}
for the downlink of wireless networks, which has been considered for
cell-free networks \cite{Ngo2017,Nguyen2017,Nayebi2017,itap&prec}.
Low-complexity conjugate beamforming (CB) has been investigated in
\cite{Ngo2017}. Zero-forcing (ZF) precoding has been studied in the
context of EE maximization in \cite{Nguyen2017}. CB and ZF precoding
have been considered with power allocation to provide uniformly good
service for all users in \cite{Nayebi2017}. \textcolor{r}{Scalable
cell-free solutions with decentralized processing and clustering
methods have been recently proposed to decrease the computational
complexity of previous techniques and facilitate their deployment
\cite{Bjornson2020a}. A minimum mean-square error (MMSE) combiner
with both centralized and decentralized implementation is presented
in \cite{Bjornson2020}. In \cite{Interdonato2020}, two distributed
precoding methods were introduced, one called local partial ZF and
local protective partial ZF. A clustering method based on a
user-centric approach has been proposed so that users are served
only by a subset of APs \cite{Buzzi2020}.}

A key problem with transmit processing techniques for cell-free
systems is how to deal with imperfect channel state information
(CSI), which can significantly degrade the system performance.
\textcolor{r}{CSI is obtained by users sending known pilot sequences
to the APs, which estimate the channel coefficients based on the
received signal. The error is caused by noise, time-varying
characteristics of the channel, which makes the obtained CSI
outdated, pilot contamination and frequency offset. In practical
systems, perfect CSI is impossible to be obtained, which calls for
robust design techniques that can deal with imperfect CSI. Most
precoders proposed in the literature consider perfect CSI, providing
unrealistic solutions to practical scenarios. Therefore, the
development of robust precoders that can deal with imperfect CSI is
of great interest.}

Robust techniques have been developed in sensor array signal
processing to mitigate the effects of uncertainties.  Although never
applied to cell-free before, robust techniques can improve the
performance of precoding and power allocation  without increasing
significantly the computational complexity. In particular, robust
techniques include diagonal loading (DL)
\cite{Li2003,Elnashar2006,Cai2015}, generalized loading
\cite{Besson2005}, worst-case optimization
\cite{Wang2013,Tajer2011,Shen2012,Guo2005,Wang2011,Shen2014} and
subspace techniques \cite{Ruan2019}.  \textcolor{r}{Recently, robust
techniques have been proposed for multiple-antenna systems. In
\cite{Medra2018}, a robust precoder with per-antenna power
constraint was considered, where the probability of outage of the
signal-to-interference-plus-noise ratio (SINR) targets is minimized
with imperfect CSI located at the base stations. Another robust
precoder was developed in \cite{Ketseoglou2019} using the ZF
criterion and user grouping, with imperfect CSI at the receivers.}

Robust precoders based on MMSE designs have been examined in
\cite{Guo2005,Tajer2011,Cai2015,Zhang2010,Wang2011,Shen2014}.
Worst-case optimization has been considered in \cite{Guo2005} by
minimizing the maximum mean-square error (MSE), taking into
consideration the channel estimation matrix and the channel
estimation error matrix. Reformulating the problem into a min-min
convex minimization problem enables the solution to be found in
closed form. Similarly, in \cite{Wang2011,Shen2014}, the same
objective function is used, with a tolerance for the channel
estimation error, yet with total power constraint.

\textcolor{r}{In this context, robust techniques have the potential
to mitigate the effects of imperfect CSI in cell-free networks,
which are particularly sensitive to the accuracy of CSI. Therefore,
the motivation of this work is to develop a robust precoder suitable
for cell-free networks. In particular, we develop an iterative
robust MMSE (RMMSE) precoder based on generalized loading along with
power allocation.} A robust precoder is calculated based on initial
parameters, used in power allocation and recalculated based on the
power allocation coefficients. MMSE channel estimates are
considered, similarly to previous works \cite{Nayebi2017}.
Additionally, optimal and uniform power allocation techniques are
devised and compared with existing CB, ZF and MMSE precoding and
power allocation techniques \cite{Nayebi2017,Ngo2017} in terms of
bit error rate (BER), sum-rate\textcolor{red}{, and per-user rate},
taking into account imperfect CSI. Moreover, analytical expressions
are derived to compute the sum rates of the proposed approaches and
their computational costs are evaluated.

\textcolor{red}{We propose a novel iterative RMMSE precoder for a
cell-free massive MIMO system scenario, which not only minimizes the
mean-square error (MSE) but also has the ability to minimize the
effect of channel estimation errors. The iterative RMMSE precoder
introduces a generalized loading technique, where a regularization
factor containing the statistics of the channel estimation error
matrix is added to the MSE objective function. We also present an
optimal and uniform power allocation strategy directly related to
the maximization of the minimum SINR, considering the proposed RMMSE
precoder. Moreover, we present a computational complexity analysis
of the proposed and existing techniques, which shows that they are
comparable in terms of computational cost. Numerical results
demonstrate that the proposed RMMSE precoder with UPA and OPA have
an improved performance compared to previous methods in terms of
BER, sum-rate and per-user rate.}

The rest of this paper is organized as follows. In Section II the
cell-free massive MIMO system model and CSI scenarios are detailed.
In Section III, an iterative RMMSE precoder with power allocation is
presented. An analysis of the sum rates and the computational
cost\textcolor{r}{s are} considered in Section IV. In Section V,
numerical results and discussions are presented, whereas in Section
VI conclusions are drawn.

\textit{Notation}: Uppercase \textcolor{r}{bold symbols denote
matrices and lowercase bold symbols denote vectors}. The
superscripts $\PC{}^*$, $\PC{}^T$, $\PC{}^H$ stand for complex
conjugate, transpose and Hermitian operations, respectively. The
expectation, trace of a matrix, real part of the argument, Euclidean
norm and Frobenius norm are denoted by $\mathbb{E}\PR{\cdot}$,
$\text{tr}\PC{\cdot}$, $\text{Re}\PC{\cdot}$, $\norm{\cdot}_2$ and
$\norm{ \cdot}_F$, \textcolor{r}{respectively}. The operator
$\text{diag} \chav{\textcolor{r}{\mathbf{V}}}$ retains the main
diagonal elements of $\textcolor{r}{\mathbf{V}}$ in a column vector.
The $D \times D$ identity matrix is $\mathbf{I}_D$ and the $D \times
E$ zero matrix is $\mathbf{0}_{D \times E}$. $x \sim
\mathcal{N}(0,\sigma^2)$ refers to a Gaussian random variable (RV)
$x$ with zero mean and variance $\sigma^2$ and $x \sim
\mathcal{CN}(0,\sigma^2)$ denotes a circularly symmetric complex
Gaussian RV $x$ with zero mean and variance $\sigma^2$.


\section{System Model}
\label{system_model}
\begin{figure}[!ht]
\centering
\includegraphics[width = 0.75\columnwidth]{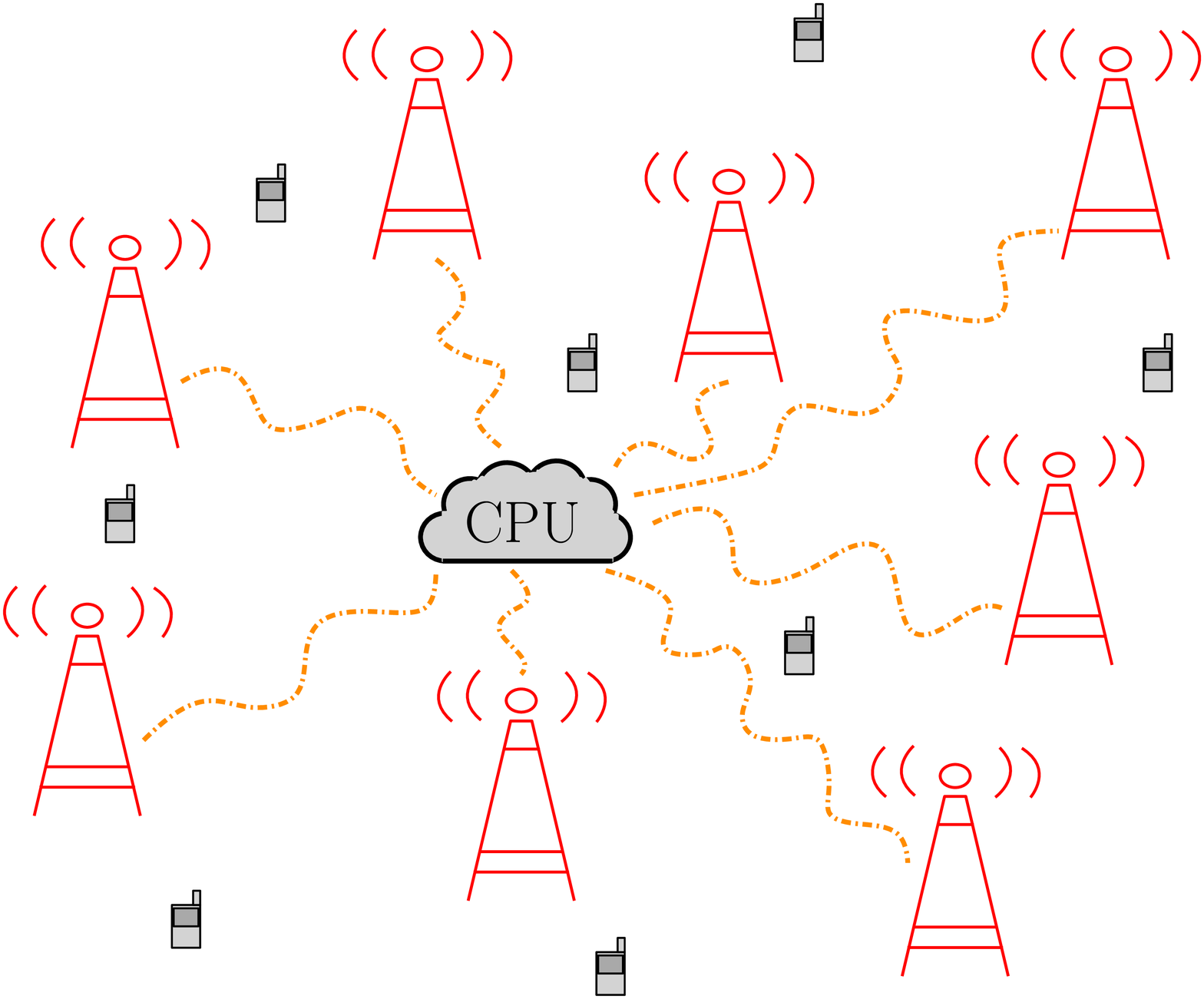}
\caption{Cell-free massive MIMO system.}\label{CF_Layout}
\end{figure}

The downlink of a cell-free massive MIMO system is considered with \textcolor{red}{$M$} randomly distributed \textcolor{red}{single-antenna} APs and $\textcolor{r}{U}$ single-antenna users, where $M >> \textcolor{r}{U}$. In this system, all APs are connected to a CPU and serve simultaneously all users, as shown in Fig.~\ref{CF_Layout}.

The channel coefficient between the $m$th \textcolor{red}{AP} and the $\textcolor{r}{u}$th user is defined as \cite{Ngo2017}
\begin{equation}
    g_{m,\textcolor{r}{u}} = \sqrt{\beta_{m,\textcolor{r}{u}}} h_{m,\textcolor{r}{u}},
    \label{def_g_mk}
\end{equation}
where $\beta_{m,\textcolor{r}{u}}$ is the large-scale fading coefficient (\textcolor{r}{due to} path loss and shadowing effects) and $h_{m,\textcolor{r}{u}} \sim \mathcal{CN}(0,1)$ is the small-scale fading coefficient, defined as independent and identically distributed (i.i.d) RVs that remain constant during a coherence interval and are independent in different coherence intervals. The large-scale fading coefficients change less frequently, being constant for several coherence intervals.  In this case, we assume that they change at least $40$ times slower than $h_{m,\textcolor{r}{u}}$ \cite{Ngo2017} and that pilot contamination is negligible \cite{Nayebi2017}.

The system employs the time division duplex (TDD) protocol, which relies on the reciprocity principle to acquire the CSI at the transmitter. Then, the CSI  is sent to the CPU, which performs precoding and power allocation to reduce the multi-user interference. By considering MMSE estimates of CSI at each AP, we define them as \cite{Nayebi2017}
\begin{equation}
    \hat{g}_{m,\textcolor{r}{u}} \sim \mathcal{CN}(0,\alpha_{m,\textcolor{r}{u}}) \text{,} \qquad \tilde{g}_{m,\textcolor{r}{u}} \sim \mathcal{CN}(0,\beta_{m,\textcolor{r}{u}} - \alpha_{m,\textcolor{r}{u}}),
\end{equation}
where $\hat{g}_{m,\textcolor{r}{u}}$ is the CSI estimate between the $m$th \textcolor{red}{AP} and the $\textcolor{r}{u}$th user and $\tilde{g}_{m,\textcolor{r}{u}}$ is the CSI error between the $m$th \textcolor{red}{AP} and the $\textcolor{r}{u}$th user. To evaluate different levels of imperfect CSI, we consider $\alpha_{m,\textcolor{r}{u}}$ as an adjustable percentage of $\beta_{m,\textcolor{r}{u}}$ ($0 \leq n \leq 1$). Thus, we have
\vspace{-0.1em}
\begin{equation}
\begin{split}
    \alpha_{m,\textcolor{r}{u}} &= n\beta_{m,\textcolor{r}{u}}\\
    \tilde{g}_{m,\textcolor{r}{u}} &= g_{m,\textcolor{r}{u}} - \hat{g}_{m,\textcolor{r}{u}}, \text{ and}\\
    \mathbb{E} \PR{\PM{\tilde{g}_{m,\textcolor{r}{u}}}^2} &= \PC{1-n} \beta_{m,\textcolor{r}{u}}.
    \end{split}
\end{equation}

\section {Proposed RMMSE Precoding and Power Allocation}
\label{mmse_ap_select_power_alloc}

\textcolor{r}{In this section, a fully digital RMMSE precoder is
derived and two power allocation techniques are introduced. We then
present an algorithm which iteratively combines the precoding and
power control solutions.}

\subsection{Proposed RMMSE Precoder}
\label{precoder_design}

In the downlink of a cell-free setting with precoding and power
allocation, the signal received by the $\textcolor{r}{u}$th user is
\begin{equation}
    y_{\textcolor{r}{u}} = \sqrt{\rho_f} \ \mathbf{g}^T_{\textcolor{r}{u}} \mathbf{P} \ \mathbf{N} \ \mathbf{s} + w_{\textcolor{r}{u}},
    \label{received_signal_u_robust}
\end{equation}
where $\rho_f$ is the maximum transmit power of each
\textcolor{red}{AP}, $\mathbf{g}_{\textcolor{r}{u}} =
\PR{g_{1,\textcolor{r}{u}},\dots,g_{M,\textcolor{r}{u}}}^T$ are the
channel coefficients for user $\textcolor{r}{u}$, $\mathbf{P} \in
\mathbb{C}^{M \times \textcolor{r}{U}}$ is the precoding matrix,
$\mathbf{N} \in \mathbb{R}_{+}^{\textcolor{r}{U} \times
\textcolor{r}{U}}$ is the power allocation diagonal matrix with
$\sqrt{\eta_1},\dots,\sqrt{\eta_{\textcolor{r}{U}}}$ on its
diagonal, $\eta_{\textcolor{r}{u}}$ is the power coefficient of user
$\textcolor{r}{u}$,  $\mathbf{s} =
[s_1,\dots,s_{\textcolor{r}{U}}]^T$ is the zero mean symbol vector,
with $\sigma_s^2 = \mathbb{E} (|s_{\textcolor{r}{u}}|^2)$,
$s_{\textcolor{r}{u}}$ is the data symbol intended for user
$\textcolor{r}{u}$, which is uncorrelated \textcolor{r}{among}
users, $w_{\textcolor{r}{u}} \sim \mathcal{CN}(0,\sigma_{w}^2)$ is
the additive noise for user $\textcolor{r}{u}$ and $\sigma_{w}^2$ is
the noise variance. \textcolor{r}{Similar to
\cite{Ngo2017,Nayebi2017}, $\mathbf{N}$ weights the symbols intended
for each user, $s_u$, by a factor $\sqrt{\eta_u}$, where $u =
1,\dots, U$. We compute $\mathbf{P}$ and $\mathbf{N}$ in an
alternating fashion to solve the precoding and power allocation
problems. The combined signal with all users is described by}

\begin{small}
\begin{equation}
\begin{split}
\mathbf{y} &= \sqrt{\rho_f} \ \mathbf{G}^T \mathbf{P} \ \mathbf{N} \mathbf{s} + \mathbf{w}\\
&= \sqrt{\rho_f} \ \PC{\hat{\mathbf{G}}+\tilde{\mathbf{G}}}^T \mathbf{P} \ \mathbf{N} \mathbf{s} + \mathbf{w}\\
&= \sqrt{\rho_f} \ \hat{\mathbf{G}}^T \mathbf{P} \ \mathbf{N} \mathbf{s} + \underbrace{\sqrt{\rho_f} \ \tilde{\mathbf{G}}^T \mathbf{P} \ \mathbf{N} \mathbf{s}}_{\boldsymbol{\Delta} } + \mathbf{w}\\
\end{split}
\end{equation}
\end{small}
\hspace{-0.2cm}where $\mathbf{G} \in \mathbb{C}^{M \times
\textcolor{r}{U}}$ is the channel matrix, \textcolor{r}{which varies
according to an ergodic stationary process}, $\hat{\mathbf{G}} \in
\mathbb{C}^{M \times \textcolor{r}{U}}$ is the CSI matrix,
$\tilde{\mathbf{G}} \in \mathbb{C}^{M \times \textcolor{r}{U}}$ is
the CSI error matrix and
$\mathbf{w}=\PR{w_1,\dots,w_{\textcolor{r}{U}}}^T$ is the noise
vector.

The design of the proposed RMMSE precoder has two main objectives:
to minimize the MSE under a total power constraint and to mitigate
the effects of CSI errors, $\mathbb{E} \PR{\lVert
\boldsymbol{\Delta}\rVert_2^2} \rightarrow 0$. Therefore, we
\textcolor{red}{add a regularization term multiplied by the scalar
$\Gamma$ to adjust the regularization effects.} The proposed RMMSE
precoder solves the following optimization:

\begin{small}
\begin{subequations} \label{rmmse_opt_problem}
\begin{equation}
\begin{split}
\chav{\mathbf{P}_{\text{RMMSE}},\mathbf{N},f_{\text{RMMSE}}} &= \text{argmin}_{\chav{\mathbf{P},\mathbf{N},f}} \mathbb{E} \PR{\norm{\mathbf{s}-f^{-1}\mathbf{y}}_2^2}\\ &\textcolor{red}{\quad + \Gamma \text{tr}\PC{\mathbf{M}\mathbf{P}\mathbf{N} \mathbf{C}_s\mathbf{N}^H\mathbf{P}^H}}
\end{split}
\end{equation}
\begin{equation}
    \rm{subject ~to~~}  \mathbb{E} \PR{\norm{\mathbf{x}}_2^2} = E_{tr}
\end{equation}
\end{subequations}
\end{small}
\hspace{-0.2cm}where \textcolor{r}{$f^{-1}$ is a normalization
factor  at the receivers, which can be interpreted as an automatic
gain control \cite{Joham2005}}, and the transmit signal is given by
\begin{equation}
    \mathbf{x} = \sqrt{\rho_f} \ \mathbf{P} \ \mathbf{N} \ \mathbf{s}.
\end{equation}

The average transmit power is described by
\begin{equation}
    \mathbb{E} \PR{\norm{\mathbf{x}}_2^2} = \rho_f \text{tr}\PC{\mathbf{P} \mathbf{N} \mathbf{C}_s \mathbf{N}^H \mathbf{P}^H} = E_{tr},
\end{equation}
where $\mathbf{C}_s = \mathbb{E} \PR{\mathbf{s}\mathbf{s}^H}$ is the symbol covariance matrix.

{The regularization factor contains }the auxiliary matrix
\begin{equation}
    \mathbf{M} = \theta \mathbb{E} \PR{\tilde{\mathbf{G}}^{*}\tilde{\mathbf{G}}^T},
\end{equation}
where $\mathbb{E} [\tilde{\mathbf{G}}^{*}\tilde{\mathbf{G}}^T]$ is a diagonal matrix with $\sum_{\textcolor{r}{u}=1}^{\textcolor{r}{U}} \PC{(1-n)\beta_{m,\textcolor{r}{u}}}$ on its $m$th diagonal element and $\theta$ is a chosen scalar. Although $\mathbf{M}$ is diagonal, the diagonal elements are not equal. The proposed \textcolor{red}{regularization} is categorized as a generalized loading \cite{Besson2005}, where the matrix is usually obtained through steering vector errors and complemented by a scalar. Here, we employ a matrix $\mathbf{M}$ obtained from the statistics of the \textcolor{r}{CSI} error matrix $\tilde{\mathbf{G}}$ and scaled by a constant $\theta$ \textcolor{red}{as an adaptation of the technique to the design of robust MMSE precoding for cell-free networks.}

By constructing the Lagrangian function with the Lagrange
multiplier, $\lambda$, setting its derivatives to zero and
considering a power allocation matrix $\mathbf{N}$, we can compute
the precoder ${\mathbf P}$ and the normalization $f$, as shown
below:
\begin{small}
\begin{equation}
\begin{split}
&\mathcal{L} \PC{\mathbf{P}, \mathbf{N}, f,\textcolor{r}{\Gamma, \lambda}}  = \mathbb{E} \PR{\norm{\mathbf{s}-f^{-1}\mathbf{y}}_2^2} + \textcolor{r}{\Gamma} \text{tr} \PC{\mathbf{M} \mathbf{P} \mathbf{N} \mathbf{C}_s \mathbf{N}^H \mathbf{P}^H}\\
&+ \lambda \PC{\rho_f \text{tr}\PC{\mathbf{P} \mathbf{N} \mathbf{C}_s \mathbf{N}^H \mathbf{P}^H} - E_{tr}} \\
&= \text{tr} \PC{\mathbf{C}_s} -  f^{-1}\sqrt{\rho_f} \text{tr} \PC{\hat{\mathbf{G}}^{T} \mathbf{P} \mathbf{N} \mathbf{C}_s} - f^{-1}\sqrt{\rho_f} \text{tr} \PC{\hat{\mathbf{G}}^{*}\mathbf{C}_s \mathbf{N}^H \mathbf{P}^H}\\
&+ f^{-2}\rho_f \text{tr} \left(\hat{\mathbf{G}}^{*} \hat{\mathbf{G}}^{T} \mathbf{P} \mathbf{N} \mathbf{C}_s \mathbf{N}^H \mathbf{P}^H\right) + f^{-2}\text{tr} \PC{\mathbf{C}_w}+ \textcolor{r}{\Gamma} \text{tr}\left(\mathbf{M} \mathbf{P} \mathbf{N} \mathbf{C}_s \right.\\
&\left. \mathbf{N}^H \mathbf{P}^H \right) + \lambda \PC{\rho_f \text{tr}\PC{\mathbf{P} \mathbf{N} \mathbf{C}_s \mathbf{N}^H \mathbf{P}^H} - E_{tr}},\\
\end{split}
\end{equation}
\end{small}
\hspace{-0.2cm}where $\mathbf{C}_w = \mathbb{E} \PR{\mathbf{w}\mathbf{w}^H}$ is the noise covariance matrix.

Using the result of the partial derivative, $\partial \text{tr}
\PC{\mathbf{B} \mathbf{X}^H}/ \partial \mathbf{X}^{*} = \mathbf{B}$,
we obtain the following expressions:
\begin{small}
\begin{equation} \label{p_robust_1}
\begin{split}
\frac{\partial \mathcal{L} \PC{\mathbf{P}, \mathbf{N}, f, \textcolor{r}{\Gamma, \lambda}}}{\partial \mathbf{P}^{*}} &= - f^{-1}\sqrt{\rho_f} \hat{\mathbf{G}}^{*}\mathbf{C}_s\mathbf{N}^H  + f^{-2}\rho_f \hat{\mathbf{G}}^{*} \hat{\mathbf{G}}^{T} \mathbf{P} \mathbf{N}\\
&\mathbf{C}_s \mathbf{N}^H + \textcolor{r}{\Gamma} \mathbf{M} \mathbf{P}\mathbf{N}\mathbf{C}_s \mathbf{N}^H  + \lambda \rho_f \mathbf{P} \mathbf{N} \mathbf{C}_s \mathbf{N}^H = 0,
\end{split}
\end{equation}
\end{small}
and
\begin{equation} \label{p_robust_2}
\begin{split}
\frac{\partial \mathcal{L} \PC{\mathbf{P}, \mathbf{N}, f, \textcolor{r}{\Gamma, \lambda}}}{\partial f} &= f^{-2}\sqrt{\rho_f} \text{tr} \PC{\hat{\mathbf{G}}^{T} \mathbf{P}\mathbf{N}\mathbf{C}_s} + f^{-2}\sqrt{\rho_f} \text{tr} \left(\hat{\mathbf{G}}^{*} \right.\\
&\left. \mathbf{C}_s \mathbf{N}^H \mathbf{P}^H \right) -2 f^{-3}\rho_f \text{tr} \left(\hat{\mathbf{G}}^{*} \hat{\mathbf{G}}^{T} \mathbf{P} \mathbf{N} \mathbf{C}_s \right.\\
& \left.\mathbf{N}^H \mathbf{P}^H \right) -2 f^{-3}\text{tr} \PC{\mathbf{C}_w} = 0.\\
\end{split}
\end{equation}

Solving for \eqref{p_robust_1}, we obtain
\begin{equation}
\mathbf{P} = \frac{f}{\sqrt{\rho_f}}
\underbrace{\PC{\hat{\mathbf{G}}^{*} \hat{\mathbf{G}}^{T} +
\frac{\textcolor{r}{\Gamma} f^2}{\rho_f} \mathbf{M} + \lambda
f^2\mathbf{I}_M}^{-1} \hat{\mathbf{G}}^{*}}_{\tilde{\mathbf{P}}}
\mathbf{N}^{-1}
\end{equation}
By using the expression in \PC{\ref{p_robust_2}}, we arrive at
\begin{equation}
    f\sqrt{\rho_f} \text{tr} \PC{\hat{\mathbf{G}}^{*}\mathbf{C}_s \mathbf{N}^H \mathbf{P}^H} = \rho_f \text{tr}
    \PC{\hat{\mathbf{G}}^{*} \hat{\mathbf{G}}^{T} \mathbf{P} \mathbf{N} \mathbf{C}_s \mathbf{N}^H \mathbf{P}^H} + \text{tr}
    \PC{\mathbf{C}_w}
    \label{p_robust_2_2}
\end{equation}
Using \eqref{p_robust_1}, we have
\begin{equation}
\begin{split}
&f\sqrt{\rho_f} \hat{\mathbf{G}}^{*}\mathbf{C}_s\mathbf{N}^H  = \rho_f \hat{\mathbf{G}}^{*} \hat{\mathbf{G}}^{T} \mathbf{P} \mathbf{N}\mathbf{C}_s\mathbf{N}^H + \textcolor{r}{\Gamma} f^2 \mathbf{M} \mathbf{P}\mathbf{N}\mathbf{C}_s\mathbf{N}^H\\
&+ \lambda f^2 \rho_f \mathbf{P} \mathbf{N}\mathbf{C}_s \mathbf{N}^H.
\label{p_robust_3}
\end{split}
\end{equation}
Pre-multiplying \textcolor{r}{on the right} by $\mathbf{P}^H$ in
\eqref{p_robust_3}, using the trace operator and considering
$\epsilon = \lambda f^2$, the expression takes the form
\begin{small}
\begin{equation}
\begin{split}
&f\sqrt{\rho_f} \text{tr}\PC{\hat{\mathbf{G}}^{*}\mathbf{C}_s\mathbf{N}^H \mathbf{P}^H} = \rho_f \text{tr}\PC{\hat{\mathbf{G}}^{*} \hat{\mathbf{G}}^{T} \mathbf{P} \mathbf{N} \mathbf{C}_s \mathbf{N}^H \mathbf{P}^H}\\
&+ \textcolor{r}{\Gamma} f^2 \text{tr}\PC{\mathbf{M} \mathbf{P} \mathbf{N} \mathbf{C}_s \mathbf{N}^H \mathbf{P}^H} + \epsilon \rho_f \text{tr}\PC{\mathbf{P} \mathbf{N} \mathbf{C}_s \mathbf{N}^H \mathbf{P}^H}.
\end{split}
\label{p_robust_3_2}
\end{equation}
\end{small}
{Equating} expression\textcolor{r}{s} \eqref{p_robust_2_2}
\textcolor{r}{and} \eqref{p_robust_3_2}, we have
\begin{small}
\begin{equation}
\begin{split}
\text{tr} \PC{\mathbf{C}_w} &= \textcolor{r}{\Gamma} f^2 \text{tr}\PC{\mathbf{M} \mathbf{P} \mathbf{N} \mathbf{C}_s \mathbf{N}^H \mathbf{P}^H} + \epsilon \rho_f \text{tr}\PC{\mathbf{P} \mathbf{N} \mathbf{C}_s \mathbf{N}^H \mathbf{P}^H} \\
\text{tr} \PC{\mathbf{C}_w} &= \textcolor{r}{\Gamma} f^2 \text{tr}\PC{\mathbf{M} \mathbf{P} \mathbf{N} \mathbf{C}_s \mathbf{N}^H \mathbf{P}^H} + \epsilon E_{tr}. \\
\end{split}
\label{epsilon_exp}
\end{equation}
\end{small}
By manipulating the expression, we obtain
\begin{small}
\begin{equation}
\epsilon = \frac{\text{tr} \PC{\mathbf{C}_w}}{E_{tr}} - \frac{\textcolor{r}{\Gamma} f_{\text{RMMSE}}^4 \text{tr}\PC{\mathbf{M} \tilde{\mathbf{P}} \mathbf{C}_s \tilde{\mathbf{P}}^H}}{\rho_f E_{tr}}\\
\end{equation}
\end{small}
\hspace{-0.2cm}where
\begin{small}
\begin{equation}
f_{\text{RMMSE}}= \sqrt{\PC{E_{tr}}/\PC{\text{tr}\PC{\tilde{\mathbf{P}}  \mathbf{C}_s \tilde{\mathbf{P}}^H}}}.
\label{exp_f_rmmse}
\end{equation}
\end{small}
Therefore, the RMMSE precoder that takes into account power allocation for cell-free networks is given by
\begin{small}
\begin{equation} \label{exp_rmmse}
\begin{split}
\mathbf{P}_{\text{RMMSE}} &= \frac{f_{\text{RMMSE}}}{\sqrt{\rho_f}} \left(\hat{\mathbf{G}}^{*} \hat{\mathbf{G}}^{T} + \frac{\text{tr} \PC{\mathbf{C}_w}}{E_{tr}}\mathbf{I}_M \right.\\
& \quad \left. + \underbrace{\frac{\textcolor{r}{\Gamma} f_{\text{RMMSE}}^2\PC{E_{tr}\mathbf{M} - f_{\text{RMMSE}}^2\text{tr}\PC{\mathbf{M} \tilde{\mathbf{P}} \mathbf{C}_s \tilde{\mathbf{P}}^H}\mathbf{I}_M}}{\rho_f E_{tr}}}_{\mathbf{F}} \right)^{-1} \\
& \quad \hat{\mathbf{G}}^{*}\mathbf{N}^{-1} = \frac{f_{\text{RMMSE}}}{\sqrt{\rho_f}}\tilde{\mathbf{P}}\mathbf{N}^{-1}.
\end{split}
\end{equation}
\end{small}
\hspace{-0.2cm}where $\text{tr} \PC{\mathbf{C}_w} = \textcolor{r}{U}
\sigma_{w}^2$. Note that if we assume perfect CSI,
\textcolor{red}{the error matrix $\tilde{\mathbf{G}}$ goes to zero,
i. e., $\tilde{\mathbf{G}} = \mathbf{0}_{M \times U}$} and the RMMSE
precoder becomes the MMSE precoder for cell-free
\textcolor{r}{networks}. \textcolor{red}{The same effect occurs if
we set $\Gamma = 0$.} Thus, the advantages of the RMMSE precoder
will be only perceived in an imperfect CSI scenario.

\subsection{Power Allocation}
 \label{power_alloc}
In this section, we introduce Optimal Power Allocation (OPA) and
Uniform Power Allocation (UPA) techniques applied to the RMMSE
precoder. The objective is to find the power allocation matrix
$\mathbf{N}$, a diagonal matrix with
$\sqrt{\eta_1},\dots,\sqrt{\eta_{\textcolor{r}{U}}}$ on its main
diagonal, which will be used to recompute the robust precoding
matrix $\mathbf{P}_{\text{RMMSE}}$ and the final power allocation
matrix $\mathbf{N}_{\text{RMMSE}}$.

\subsubsection{Optimal Power Allocation (OPA)}
\label{opt_power_alloc}

{The chosen power allocation technique is optimal in terms of
ensuring fairness among all users. The main point of this approach
is to improve the performance of the user with the lowest SINR,
meaning that all users will have a least a certain quality of
service (QoS).} The max-min fairness power allocation problem with
{AP} power constraint for the RMMSE precoder can be formulated as

\begin{subequations}
\begin{align}
&\max_{\boldsymbol{\eta}} \min_{\textcolor{r}{u}} \text{SINR}_{\textcolor{r}{u}} \PC{\boldsymbol{\eta}} \label{sinr_mmse_robust_complete}\\
&\text{s.t.} \sum_{i=1}^{\textcolor{r}{U}} \eta_{i} \delta_{m,i} \leq 1, m=1,\dots,M,
\end{align}
\end{subequations}
where $\text{SINR}_{\textcolor{r}{u}}\PC{\boldsymbol{\eta}}$ denotes the SINR of user $\textcolor{r}{u}$ \textcolor{r}{as a function of $\boldsymbol{\eta}$} and can be expressed as
\textcolor{r}{
\begin{small}
\begin{equation}
    \text{SINR}_{\textcolor{r}{u}}\PC{\boldsymbol{\eta}} = \frac{\mathbb{E} [|C_1\PC{\boldsymbol{\eta}}|^2]}{\PC{\sigma_w^2 + \sum_{i=1,i \neq \textcolor{r}{u}}^{\textcolor{r}{U}}\mathbb{E} [|C_{2,i}\PC{\boldsymbol{\eta}}|^2]+\mathbb{E} [|C_3\PC{\boldsymbol{\eta}}|^2]}}.
\end{equation}
\end{small}
}
The term $\textcolor{r}{C_1\PC{\boldsymbol{\eta}}}$ is the desired
signal, $\sigma_w^2$ is the noise variance,
$\textcolor{r}{C_{2,i}\PC{\boldsymbol{\eta}}}$ is the interference
caused by user $i \ \textrm{for} \ i \neq \textcolor{r}{u}, i =
1,\dots,\textcolor{r}{U}$,
$\textcolor{r}{C_3\PC{\boldsymbol{\eta}}}$ is the CSI error,
$\boldsymbol{\delta}_m = \mathrm{diag} \chav{ \mathbb{E}
\PR{\mathbf{p}_{m}^T \mathbf{p}_{m}^{*}}}, m=1,\dots,M$,
$\mathbf{p}_{m} = \PR{p_{m,1},\dots,p_{m,\textcolor{r}{U}}}$ is the
$m$th row of the precoder $\mathbf{P}_{\text{RMMSE}}$ and
$\delta_{m,i}$ is the $i$th element of vector
$\boldsymbol{\delta}_m$.

The power allocation problem can be expressed in an epigraph form as
\textcolor{r}{\cite{Ngo2017,Nayebi2017,StephenBoyd2019}}
\begin{subequations}\label{epigraph_opa_robust}
\begin{align}
& \text{find } \boldsymbol{\eta} \\
&\text{s.t.} \ \text{SINR}_{\textcolor{r}{u}} \PC{\boldsymbol{\eta}} \geq t, \ \textcolor{r}{u}=1,\dots,\textcolor{r}{U},\\
& \qquad \sum_{i=1}^{\textcolor{r}{U}} \eta_{i} \delta_{m,i} \leq 1, \ m=1,\dots,M,
\end{align}
\end{subequations}
where $t = \frac{t_b + t_e}{2}$ is the midpoint of a chosen interval $(t_b,t_e)$.

{The problem in (24) is solved by addressing one feasibility problem
at each step. First we assume that the problem is feasible. We set
up the interval, $[t_b,t_e]$ known to contain the optimal value
$t^{*}$. Next, the convex feasibility problem is solved at its
midpoint $t = \frac{t_b + t_e}{2}$, to decide whether the problem is
feasible or not. If the problem is feasible, the upper interval is
chosen, meaning that $t_b = t$. If, however, the problem is
infeasible, the lower interval will now be evaluated, translating to
$t_e = t$. The new interval will be half of the previous and the
algorithm will be repeated until the resulting interval is smaller
than a certain tolerance $\varepsilon$, $|t_e - t_b| < \varepsilon$,
\cite{StephenBoyd2019}}.

\subsubsection{Uniform Power Allocation (UPA)}
\label{unif_power_alloc}

We also present an alternative to the OPA scheme, based on
\cite{Nayebi2017}. \textcolor{r}{In scenarios} where a certain {AP}
$m$ transmits with full power and all $\eta_{{u}}$, for
$\textcolor{r}{u}=1,\dots,\textcolor{r}{U}$ are equal, we have
\begin{align}
\eta_{\textcolor{r}{u}} &= 1 / \PC{\max_{m} \sum_{i=1}^{\textcolor{r}{U}} \delta_{m,i}}, \ \textcolor{r}{u}=1,\dots,\textcolor{r}{U}, \label{upa_opt_robust}
\end{align}
where $\delta_{m,i}$ is the $i$th element of vector
$\boldsymbol{\delta}_m$. Although \eqref{upa_opt_robust} is a
suboptimal solution, it is less complex and a cost-effective
alternative to show the benefits of the RMMSE precoder. {Next, in
Algorithm~\ref{alg_mmse_robust_total_power} we show how to combine
the RMMSE precoder with power allocation. Each of the loops performs
two iterations such that $\text{ITER}_{\text{prec}} =
\text{ITER}_{\text{pa}} = 2$. }

{We initialize the algorithm with the precoding and power allocation
matrix $\tilde{\mathbf{P}}_{\text{MMSE}}$ and
$\mathbf{N}_{\text{MMSE}}$ obtained through the MMSE criterion with
$\textcolor{r}{\Gamma} = 0$. After the MMSE precoder is obtained, we
recursively calculate our initial  $f_{\text{RMMSE}}$ and
$\tilde{\mathbf{P}}$, described in \eqref{exp_f_rmmse} and
\eqref{exp_rmmse}, respectively. Then, we enter the power allocation
loop, where in the first iteration, $\mathbf{P}_{\text{RMMSE}}$ is
calculated considering $\mathbf{N}[1]= \mathbf{N}_{\text{MMSE}}$.
The subsequent step is to find a new power allocation matrix
$\mathbf{N}$ based on the calculated precoder,
$\mathbf{P}_{\text{RMMSE}}$. Then, the next precoding matrix will be
computed with the new $\mathbf{N}$. The final step of the loop is to
recalculate $\mathbf{N}$ with the last RMMSE precoder found. Note
that the final power allocation matrix $\mathbf{N}_{\text{RMMSE}}$
satisfies the power constraints and is different from the
intermediate $\mathbf{N}$ present in the precoding expression.
Therefore they will not cancel each other. The scalar $\theta$ is
set as -1  to ensure that the diagonal elements of $\mathbf{F}$ are
always positive. Then ${\Gamma}$ is chosen to maximize the SINR of
the system.}

\begin{algorithm}[ht!]
\textcolor{r}{
\caption{Proposed Iterative RMMSE Precoding}
\label{alg_mmse_robust_total_power}
\begin{algorithmic}[1]
\State Find $f_{\text{MMSE}}$, $\tilde{\mathbf{P}}_{\text{MMSE}}$ and $\mathbf{N}_{\text{MMSE}}$, by considering $\textcolor{r}{\Gamma} = 0$
\State Initialize $f_{\text{RMMSE}}[1]=f_{\text{MMSE}}$, $\tilde{\mathbf{P}}[1] = \tilde{\mathbf{P}}_{\text{MMSE}}$, $\theta = -1$, $\mathbf{N}[1]= \mathbf{N}_{\text{MMSE}}$, $\text{ITER}_{\text{prec}}$ (number of iterations for the precoder), $\text{ITER}_{\text{pa}}$ (number of iterations for power allocation).
\State{\textbf{For} i=1:$\text{ITER}_{\text{prec}}$}
\Statex {Update} $\textcolor{r}{\Gamma}$, $\tilde{\mathbf{P}}$ {and} $f_{\text{RMMSE}}$:
\State Calculate $\textcolor{r}{\Gamma}[i+1]$ to optimize the SINR.
\State Calculate $\tilde{\mathbf{P}}[i+1] \gets \eqref{exp_rmmse}$
\State Calculate $f_{\text{RMMSE}}[i+1] \gets \eqref{exp_f_rmmse} $
\State{\textbf{end for}}
\State Obtain $\tilde{\mathbf{P}} = \tilde{\mathbf{P}}[i+1]$ and $f_{\text{RMMSE}} = f_{\text{RMMSE}}[i+1]$
\State{\textbf{For} j=1:$\text{ITER}_{\text{pa}}$}
\State Calculate $\mathbf{P}_{\text{RMMSE}}[j] \gets \eqref{exp_rmmse}$
\State Calculate $\mathbf{N}[j+1] \gets$ \eqref{epigraph_opa_robust} or \eqref{upa_opt_robust}(with fixed $\mathbf{P}_{\text{RMMSE}}[j]$)
\State{\textbf{end for}}
\State Obtain $\mathbf{P}_{\text{RMMSE}} = \mathbf{P}_{\text{RMMSE}}[j]$ and $\mathbf{N}_{\text{RMMSE}} = \mathbf{N}[j+1]$.
\end{algorithmic}}
\end{algorithm}

\section{{Sum-rate and Complexity Analysis}}
\label{sum_rate_analysis}

In this section, we present a sum-rate analysis of the proposed
techniques along with the computational complexity of the proposed
and existing algorithms.

\subsection{Sum-Rate}

Expanding expression \eqref{received_signal_u_robust}, we obtain:
\begin{equation}
\begin{split}
y_{{u}} &= \sqrt{\rho_f} \ \mathbf{g}^{T}_{\textcolor{r}{u}} \mathbf{P}_{\text{RMMSE}} \ \mathbf{N}_{\text{RMMSE}} \ \mathbf{s} + w_{\textcolor{r}{u}} \\
&= \underbrace{\sqrt{\rho_f} \ \hat{\mathbf{g}}^{T}_{\textcolor{r}{u}} \mathbf{P}_{\text{RMMSE}} \mathbf{N}_{\text{RMMSE}}  \ \mathbf{s}}_{\text{desired signal + interference}} + \underbrace{\sqrt{\rho_f} \ \tilde{\mathbf{g}}^{T}_{\textcolor{r}{u}} \mathbf{P}_{\text{RMMSE}} \mathbf{N}_{\text{RMMSE}} \ \mathbf{s}}_{\text{CSI error}}\\
& \qquad + \ w_{\textcolor{r}{u}}, \vspace{-0.5em}
\end{split}
\end{equation}
where $\hat{\mathbf{g}}_{\textcolor{r}{u}} =
\PR{\hat{g}_{1,\textcolor{r}{u}},\dots,\hat{g}_{M,\textcolor{r}{u}}}^T$
is the CSI vector for user $\textcolor{r}{u}$ and
$\tilde{\mathbf{g}}_{\textcolor{r}{u}} =
\PR{\tilde{g}_{1,\textcolor{r}{u}},\dots,\tilde{g}_{M,\textcolor{r}{u}}}^T$
is the CSI error vector for user $\textcolor{r}{u}$.

Assuming Gaussian signalling, \textcolor{r}{and based on the worst-case uncorrelated additive noise,} the achievable rate of the $\textcolor{r}{u}$th user with the iterative RMMSE precoder is equal to \cite{Ngo2017,Nayebi2017}
\begin{equation}
R_{\textcolor{r}{u},\text{RMMSE}} = \log_2 (1+ \text{SINR}_{\textcolor{r}{u},\text{RMMSE}})\textcolor{r}{.}
\label{r_u_rmmse}
\end{equation}
\textcolor{r}{The} sum-rate \textcolor{r}{expression} is \textcolor{r}{given by} $R_{\text{RMMSE}} = \sum_{\textcolor{r}{u}=1}^{\textcolor{r}{U}} R_{\textcolor{r}{u},\text{RMMSE}}$\textcolor{r}{.}
The $\text{SINR}_{\textcolor{r}{u},\text{RMMSE}}$ is described by
\begin{small}
\begin{equation}
\label{sinr_mmse_opt}
\text{SINR}_{\textcolor{r}{u},\text{RMMSE}} = \frac{\rho_f \eta_{\textcolor{r}{u}} \psi_{\textcolor{r}{u}}}{\sigma_{w}^2 + \rho_f \sum_{i=1,i \neq \textcolor{r}{u}}^{\textcolor{r}{U}} \eta_i \phi_{\textcolor{r}{u},i} + \rho_f \sum_{i=1}^{\textcolor{r}{U}} \eta_i \gamma_{\textcolor{r}{u},i}},
\end{equation}
\end{small}
\hspace{-0.2cm}where $\psi_{\textcolor{r}{u}} = \mathbf{p}_{\textcolor{r}{u}}^H \hat{\mathbf{g}}^{*}_{\textcolor{r}{u}} \hat{\mathbf{g}}^{T}_{\textcolor{r}{u}} \mathbf{p}_{\textcolor{r}{u}}$,
$\phi_{\textcolor{r}{u},i} = \mathbf{p}_{i}^H \hat{\mathbf{g}}^{*}_{\textcolor{r}{u}} \hat{\mathbf{g}}^{T}_{\textcolor{r}{u}} \mathbf{p}_{i}$, $i \neq \textcolor{r}{u}$, $i = 1,\dots,\textcolor{r}{U}$, $\mathbf{p}_{\textcolor{r}{u}} = \PR{p_{1,\textcolor{r}{u}},\dots,p_{M,\textcolor{r}{u}}}^T$ is the column $\textcolor{r}{u}$ of matrix $\mathbf{P}_{\text{RMMSE}}$,
$\boldsymbol{\gamma}_{\textcolor{r}{u}} = \text{diag} \chav{\mathbf{P}_{\text{RMMSE}}^H\mathbb{E} \PR{\tilde{\mathbf{g}}^{*}_{\textcolor{r}{u}} \tilde{\mathbf{g}}^{T}_{\textcolor{r}{u}}} \mathbf{P}_{\text{RMMSE}}}$, $\gamma_{\textcolor{r}{u},i}$ is the $i$th element of vector $\boldsymbol{\gamma}_{\textcolor{r}{u}}$, and  $\mathbb{E} \PR{\tilde{\mathbf{g}}^{*}_{\textcolor{r}{u}} \tilde{\mathbf{g}}^{T}_{\textcolor{r}{u}}}$ is a diagonal matrix with $\PC{(1-n)\beta_{m,\textcolor{r}{u}}}$ on its $m$th diagonal element.

\textcolor{r}{To arrive at \eqref{sinr_mmse_opt}, we assumed that
the symbols, noise and channel coefficients are mutually
independent. With fading, the ergodic rate is taken by the
expectation over all channel realizations, which can be obtained
through a rate average, considering that sufficiently long codewords
are transmitted.} Since in
$\text{SINR}_{\textcolor{r}{u},\text{RMMSE}}\PC{\boldsymbol{\eta}}$
the numerator and denominator are linear functions of
$\boldsymbol{\eta}$, the expression is a quasilinear function,
enabling us to use the bisection method \cite{StephenBoyd2019}.

\subsection{Computational Complexity}
\label{comp_complexity}

We evaluate \textcolor{r}{in this section} the computational
complexity of the \textcolor{r}{proposed} and existing methods. As
shown in Table~\ref{table_1}, the complexity of the RMMSE precoder
is comparable to the MMSE \textcolor{r}{precoder (when $\Gamma=0$)}
and the ZF precoder from \cite{Nayebi2017}. The CB precoder
\cite{Ngo2017,Nayebi2017} \textcolor{r}{has inferior performance and
much lower computational complexity as compared to the RMMSE, MMSE
and ZF precoders}. If $M^3 > T_{\text{OPA}}\textcolor{r}{U}^{3.5}$,
the computational cost of RMMSE\textcolor{r}{, MMSE and ZF with} OPA
will be $\mathcal{O}\PC{M^3}$. Depending on the number of iterations
of the bisection method, $T_{\text{OPA}}$, the OPA scheme may
prevail when applied to all precoders. In contrast, if UPA is
applied, it will not affect the complexity of the RMMSE, MMSE and ZF
techniques. We conclude that the proposed RMMSE precoder has
comparable computational cost to previous methods but can outperform
existing precoders in the presence of imperfect CSI, as shown in
Section \ref{sec_num_results}.

\begin{table}[ht]
\caption{Computational Complexity}
\vspace{-0.5em}
\centering
\resizebox{0.45\textwidth}{!}{%
    \begin{tabular}{|c|c|c|}
\hline
\multirow{4}{*}{\makecell{Precoding \\ + \\ Power Allocation}} & RMMSE Precoder& $\mathcal{O}\PC{M^3}$\\
\cline{2-3}
& MMSE Precoder & $\mathcal{O}\PC{M^3}$\\
\cline{2-3}
 & ZF Precoder & $\mathcal{O}\PC{M^3}$\\
\cline{2-3}
 & CB Precoder \cite{Ngo2017,Nayebi2017}& $\mathcal{O}\PC{M\textcolor{r}{U}}$\\
\hline
\multirow{4}{*}{SINR Computation} & RMSE Precoder& $\mathcal{O}\PC{M^2\textcolor{r}{U}^2}$\\
\cline{2-3}
& MMSE Precoder & $\mathcal{O}\PC{M^2\textcolor{r}{U}^2}$\\
\cline{2-3}
& ZF Precoder & $\mathcal{O}\PC{M^2\textcolor{r}{U}^2}$\\
\cline{2-3}
& CB Precoder & $\mathcal{O}\PC{M\textcolor{r}{U}^2}$\\
\hline
\multirow{2}{*}{Power Allocation} &  OPA & $\mathcal{O}\PC{T_{\text{OPA}}\textcolor{r}{U}^{3.5}}$\\
\cline{2-3}
& UPA & $\mathcal{O}\PC{M\textcolor{r}{U}^2}$\\
\hline
\end{tabular}}
\label{table_1}
\end{table}

\section {Numerical Results} \label{sec_num_results}

In this section, we compare the RMMSE precoder to the CB and ZF
precoders and most importantly, to the MMSE scheme. Note that all
simulations consider imperfect CSI since under perfect CSI the RMMSE
precoder converges to the MMSE precoder for cell-free. In all
experiments, we performed \textcolor{r}{500} channel realizations
and assumed $\sigma_s^2 = 1$.

\textcolor{r}{We consider the APs and users to be} uniformly distributed within an area of $1 \text{ km}^2$. The large-scale fading coefficients from (\ref{def_g_mk}) are modeled by
\begin{equation}
    \beta_{m,\textcolor{r}{u}} = \textcolor{r}{\Upsilon}_{m,\textcolor{r}{u}} \cdot 10^{\frac{\sigma_{sh} z_{m,\textcolor{r}{u}}}{10}} \vspace{-0.5em},
\end{equation}
where $\textcolor{r}{\Upsilon}$ is the path loss and $10^{\frac{\sigma_{sh} z_{m,\textcolor{r}{u}}}{10}}$ refers to the shadow fading with standard deviation $\sigma_{sh} = 8$ dB and $z_{m,\textcolor{r}{u}} \sim \mathcal{N} \PC{0,1}$.
The path loss is based on a three-slope model \cite{Tang2001}, in dB, defined as
\begin{small}
\begin{equation}
\textcolor{r}{\Upsilon}_{m,\textcolor{r}{u}} =
\begin{cases}
    -L-35 \log_{10} \PC{d_{m,\textcolor{r}{u}}}, \text{ if } d_{m,\textcolor{r}{u}} > d_1\\
    -L-15\log_{10} \PC{d_1} - 20\log_{10}\PC{d_{m,\textcolor{r}{u}}},\\
    \qquad \qquad \qquad \text{ if } d_{0} < d_{m,\textcolor{r}{u}} \leq d_1\\
    -L-15\log_{10} \PC{d_1} - 20\log_{10}\PC{d_0} , \text{ if } d_{m,\textcolor{r}{u}} \leq d_0\\
    \end{cases}
\end{equation}
\end{small}
\hspace{-0.2cm}where
\begin{small}
\begin{equation}\begin{split}
L &\triangleq 46.3 + 33.9 \log_{10} \PC{\textcolor{r}{\varrho}} - 13.82 \log_{10} \PC{h_{\text{AP}}}\\
& \qquad - \PC{1.1 \log_{10} \PC{\textcolor{r}{\varrho}} - 0.7}h_{\text{u}} + \PC{1.56 \log_{10} \PC{\textcolor{r}{\varrho}} - 0.8},
\end{split}
\end{equation}
\end{small}
$d_{m,\textcolor{r}{u}}$ is the distance between the $m$th \textcolor{red}{AP} and the $\textcolor{r}{u}$th user, $d_1 = 50$ m, $d_0 = 10$ m, $\textcolor{r}{\varrho} = 1900$ MHz is the carrier frequency in MHz, $h_{\text{AP}} = 15$ m is the AP antenna height in meters and $h_{\text{u}} = 1.65$ m is the user antenna height in meters, as in \cite{Ngo2017}. When $d_{m,\textcolor{r}{u}} \leq d_1$ there is no shadowing.

We consider strong path loss, which is typical of cell-free systems, and define \textcolor{r}{the signal-to-noise ratio (SNR) as
\begin{small}
\begin{equation}
 \text{SNR} = \frac{\rho_f \mathbb{E} [ || \hat{\mathbf{G}}||_F^2 ]}{\text{tr}\PC{\mathbf{C}_w}} = \frac{\rho_f \text{tr} (\hat{\mathbf{G}} \hat{\mathbf{G}}^H)}{\textcolor{r}{U} \sigma_{w}^2},
 \label{exp_rho_f}
\end{equation}
\end{small}}
\hspace{-0.1cm}where $\sigma_{w}^2 = T_0 \times k_B \times B \times NF \text{(W)}$, $T_0 = 290$ (Kelvin) is the noise temperature, $k_B = 1.381 \times 10^{-23}$ (Joule per Kelvin) is the Boltzmann constant, $B = 20$ MHz is the bandwidth and $NF = 9$ dB is the noise figure. \textcolor{r}{In \eqref{exp_rho_f}, we fix $\mathbf{C}_w$ and $\hat{\mathbf{G}}$ and vary $\rho_f$ according to the change in SNR values, which will be made throughout the experiments.}

\begin{figure}[!ht]
\centering
\includegraphics[width=0.8\columnwidth]{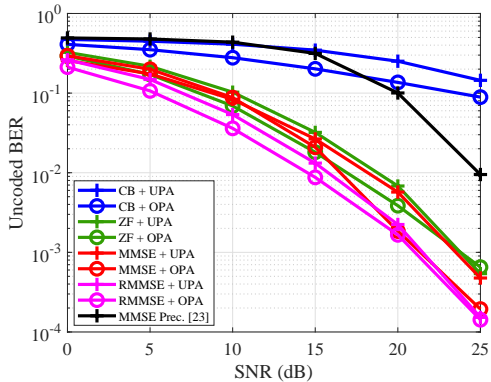} \vspace{-0.5em}
\caption{BER vs. SNR with $\textcolor{red}{M} = 96$, $\textcolor{r}{U} = 8$, $n = 0.99$, \textcolor{r}{$500$} channel realizations, $100$ symbols per packet, $E_{tr} = M\rho_f$.}\label{BER_RMMSE_OPA_2}
\end{figure}

In the first example, shown in
Fig.~\ref{BER_RMMSE_OPA_2}\textcolor{r}{,} we compare the strategies
in terms of BER vs. SNR with UPA and OPA. We assume imperfect CSI
\textcolor{r}{with} $n = 0.99$ \textcolor{r}{($1\%$)}, quadrature
phase shift keying (QPSK) modulation and \textcolor{r}{packets with
100 symbols (or 200 bits).} This scenario assumes
$\textcolor{red}{M} = 96$ \textcolor{red}{single-antenna} APs and
$\textcolor{r}{U} = 8$ users.

The gains of the RMMSE precoder are substantial over the MMSE
precoder of \cite{Joham2005}. \textcolor{r}{In addition, RMMSE with
OPA can outperform MMSE with OPA and ZF with UPA by up to $3$ dB and
$4.6$, respectively, at BER = $0.02$. Note that the RMMSE precoder
with OPA attains a consistent gain between $1.1$ and $1.7$ dB when
compared to UPA, up to SNR = 20 dB.}

\begin{table}[ht]
\caption{BER with UPA at SNR = 25 dB}
\centering
\resizebox{0.35\textwidth}{!}{%
    \begin{tabular}{|c|c|c|c|}
\hline
 & $n = 0.99$ & $n = 0.95$ & $n = 0.9$\\
 \hline
 RMMSE & $\textcolor{r}{1.25} \times 10^{-4}$ & $\textcolor{r}{3.49} \times 10^{-4}$ & $\textcolor{r}{1.6} \times 10^{-3}$  \\
\hline
 MMSE & $\textcolor{r}{3.64 \times 10^{-4}}$ & $\textcolor{r}{1.1} \times 10^{-3}$ & $\textcolor{r}{3.4} \times 10^{-3}$ \\
\hline
 ZF & $\textcolor{r}{4.3 \times 10^{-4}}$ & $\textcolor{r}{1.2} \times 10^{-3}$ & $\textcolor{r}{3.5} \times 10^{-3}$ \\
\hline
\end{tabular}}
\label{table_2}
\end{table}

In Table~\ref{table_2} we assess the BER of different precoding
schemes at SNR = 25 dB. To this end, we vary the level of CSI
imperfection, by modifying $n$, while the other parameters remain
the same as in the first example.

\textcolor{r}{Notice that for $n = 0.99$ the difference between the
BER of the RMMSE and the MMSE precoder is  $2.39 \times 10^{-4}$.
When $n$ is reduced to $95\%$, the gap is increased to $7.56 \times
10^{-4}$. For $n = 0.9$, the BER of the RMMSE precoder has a
difference of $1.8 \times 10^{-3}$, compared to that from the MMSE
precoder.}

\begin{figure}[!ht]
\centering
\includegraphics[width=0.8\columnwidth]{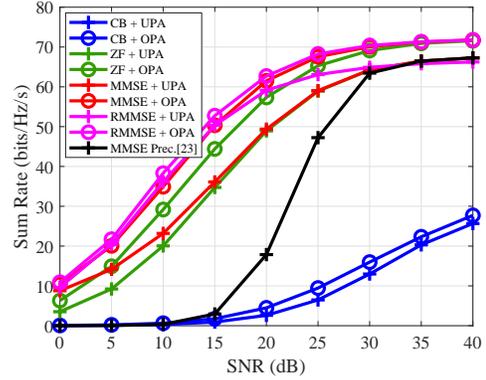}
\caption{Sum-rate vs. SNR with $\textcolor{red}{M}$ = 128,
$\textcolor{r}{U} = 16$, $n = 0.9$, \textcolor{r}{$500$} channel
realizations and $E_{tr} = M\rho_f$.}\label{SR_RMMSE_OPA}
\end{figure}

The last {experiment} compares the analyzed techniques in terms of
sum-rate vs. SNR {and cumulative distribution function (CDF) vs.
per-user rate} with $n = 0.9$ ($10\%$). Moreover, we enlarge the
system and \textcolor{r}{employ} $\textcolor{red}{M} = 128$
{single-antenna} APs, and $\textcolor{r}{U} = 16$ users. {In
Fig.~\ref{SR_RMMSE_OPA},} when OPA is \textcolor{r}{used with}
precoding, all rates are improved as compared to UPA.
\textcolor{r}{For the} RMMSE \textcolor{r}{precoder the use of} OPA
\textcolor{r}{instead of} UPA \textcolor{r}{results in} a gain of
\textcolor{red}{13\%} at SNR = 0 dB. Significant gains are
\textcolor{r}{also} obtained for \textcolor{r}{the} RMMSE
\textcolor{r}{precoder} with OPA over MMSE with UPA, which is around
\textcolor{red}{65\%} \textcolor{r}{at} SNR = \textcolor{red}{10}
dB, and for RMMSE with OPA over ZF with OPA, which can be up to
\textcolor{red}{72\%}.

\begin{figure}[!ht]
\centering
\includegraphics[width=0.8\columnwidth]{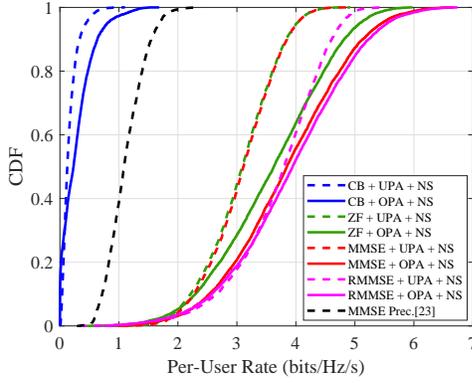}
\caption{{CDF vs. per-user rate with $M$ = 128, $U = 16$, $n = 0.9$,
$500$ channel realizations, SNR = $20$ dB and $E_{tr} =
M\rho_f$.}}\label{figure_cdf}
\end{figure}

{In terms of per-user rate, we can see in Fig.~\ref{figure_cdf} that
RMMSE+OPA has a $95\%$ probability of having values smaller or equal
to $5.5$. On the other hand, RMMSE+UPA and MMSE+UPA, have a $95\%$
chance of obtaining values smaller or equal to $4.78$ and $4.0$,
respectively. At last, ZF+OPA has a $95\%$ probability of having
values smaller or equal to $5.08$.} Future work might focus on
detection and decoding techniques for cell-free networks
\cite{deLamare2003,itic,spa,cai2009,jiols,jiomimo,mfdf,mbdf,did,rrmser,jidf,bfidd,1bitidd,detection_review,aaidd,listmtc,dynmtc,mwc,dynovs,dopeg,memd,vfap}.

\section{Conclusions}

{In this work, we have developed a novel iterative RMMSE precoder
based on generalized loading to mitigate the effects of imperfect
CSI in cell-free networks. We devised optimal and uniform power
allocation techniques based on the maximization of the minimum SINR,
taking the RMMSE precoder into account. We also performed a sum-rate
analysis and evaluated the computational cost of the proposed and
existing methods, which shows that the proposed approaches have
comparable cost. Numerical results illustrated that the proposed
techniques have improved performance in terms of BER, sum-rate and
per-user rate in imperfect CSI scenarios.}

\ifCLASSOPTIONcaptionsoff
  \newpage
\fi




\vspace{-0.2em}
\bibliographystyle{IEEEtran}
\bibliography{IEEEabrv,Bibliography}

\begin{thebibliography}{10}
\providecommand{\url}[1]{#1}
\csname url@rmstyle\endcsname
\providecommand{\newblock}{\relax}
\providecommand{\bibinfo}[2]{#2}
\providecommand\BIBentrySTDinterwordspacing{\spaceskip=0pt\relax}
\providecommand\BIBentryALTinterwordstretchfactor{4}
\providecommand\BIBentryALTinterwordspacing{\spaceskip=\fontdimen2\font plus
\BIBentryALTinterwordstretchfactor\fontdimen3\font minus
  \fontdimen4\font\relax}
\providecommand\BIBforeignlanguage[2]{{%
\expandafter\ifx\csname l@#1\endcsname\relax
\typeout{** WARNING: IEEEtran.bst: No hyphenation pattern has been}%
\typeout{** loaded for the language `#1'. Using the pattern for}%
\typeout{** the default language instead.}%
\else
\language=\csname l@#1\endcsname
\fi
#2}}

\bibitem{mmimo}
R.~C. {de Lamare}, ``Massive mimo systems: Signal processing challenges and
  future trends,'' \emph{URSI Radio Science Bulletin}, vol. 2013, no. 347, pp.
  8--20, 2013.

\bibitem{wence}
W.~{Zhang}, H.~{Ren}, C.~{Pan}, M.~{Chen}, R.~C. {de Lamare}, B.~{Du}, and
  J.~{Dai}, ``Large-scale antenna systems with ul/dl hardware mismatch:
  Achievable rates analysis and calibration,'' \emph{IEEE Transactions on
  Communications}, vol.~63, no.~4, pp. 1216--1229, 2015.

\bibitem{Marzetta2016}
T.~L. Marzetta, E.~G. Larsson, H.~Yang, and H.~Q. Ngo, \emph{Fundamentals of
  Massive {MIMO}}.\hskip 1em plus 0.5em minus 0.4em\relax Cambridge University
  Press, nov 2016.

\bibitem{Ngo2017}
H.~Q. Ngo, A.~Ashikhmin, H.~Yang, E.~G. Larsson, and T.~L. Marzetta,
  ``Cell-free massive {MIMO} versus small cells,'' \emph{{IEEE} Transactions on
  Wireless Communications}, vol.~16, no.~3, pp. 1834--1850, mar 2017.

\bibitem{Nguyen2017}
L.~D. Nguyen, T.~Q. Duong, H.~Q. Ngo, and K.~Tourki, ``Energy efficiency in
  cell-free massive {MIMO} with zero-forcing precoding design,'' \emph{{IEEE}
  Communications Letters}, vol.~21, no.~8, pp. 1871--1874, aug 2017.

\bibitem{Spencer1}
Q.~H. {Spencer}, A.~L. {Swindlehurst}, and M.~{Haardt}, ``Zero-forcing methods
  for downlink spatial multiplexing in multiuser mimo channels,'' \emph{IEEE
  Transactions on Signal Processing}, vol.~52, no.~2, pp. 461--471, 2004.

\bibitem{Stankovic}
V.~{Stankovic} and M.~{Haardt}, ``Generalized design of multi-user mimo
  precoding matrices,'' \emph{IEEE Transactions on Wireless Communications},
  vol.~7, no.~3, pp. 953--961, 2008.

\bibitem{Sung}
H.~{Sung}, S.~. {Lee}, and I.~{Lee}, ``Generalized channel inversion methods
  for multiuser mimo systems,'' \emph{IEEE Transactions on Communications},
  vol.~57, no.~11, pp. 3489--3499, 2009.

\bibitem{Zu_CL}
K.~{Zu} and R.~C. de~{Lamare}, ``Low-complexity lattice reduction-aided
  regularized block diagonalization for mu-mimo systems,'' \emph{IEEE
  Communications Letters}, vol.~16, no.~6, pp. 925--928, 2012.

\bibitem{Zu}
K.~{Zu}, R.~C. {de Lamare}, and M.~{Haardt}, ``Generalized design of
  low-complexity block diagonalization type precoding algorithms for multiuser
  mimo systems,'' \emph{IEEE Transactions on Communications}, vol.~61, no.~10,
  pp. 4232--4242, 2013.

\bibitem{wlbd}
W.~{Zhang}, R.~C. {de Lamare}, C.~{Pan}, M.~{Chen}, J.~{Dai}, B.~{Wu}, and
  X.~{Bao}, ``Widely linear precoding for large-scale mimo with iqi: Algorithms
  and performance analysis,'' \emph{IEEE Transactions on Wireless
  Communications}, vol.~16, no.~5, pp. 3298--3312, 2017.

\bibitem{rsbd}
A.~R. {Flores}, R.~C. {de Lamare}, and B.~{Clerckx}, ``Linear precoding and
  stream combining for rate splitting in multiuser mimo systems,'' \emph{IEEE
  Communications Letters}, vol.~24, no.~4, pp. 890--894, 2020.

\bibitem{rmmse}
Y.~{Cai}, R.~C. {de Lamare}, L.~{Yang}, and M.~{Zhao}, ``Robust mmse precoding
  based on switched relaying and side information for multiuser mimo relay
  systems,'' \emph{IEEE Transactions on Vehicular Technology}, vol.~64, no.~12,
  pp. 5677--5687, 2015.

\bibitem{wlbf}
N.~{Song}, W.~U. {Alokozai}, R.~C. {de Lamare}, and M.~{Haardt}, ``Adaptive
  widely linear reduced-rank beamforming based on joint iterative
  optimization,'' \emph{IEEE Signal Processing Letters}, vol.~21, no.~3, pp.
  265--269, 2014.

\bibitem{locsme}
H.~{Ruan} and R.~C. {de Lamare}, ``Robust adaptive beamforming using a
  low-complexity shrinkage-based mismatch estimation algorithm,'' \emph{IEEE
  Signal Processing Letters}, vol.~21, no.~1, pp. 60--64, 2014.

\bibitem{okspme}
H.~Ruan and R.~C. de~Lamare, ``Robust adaptive beamforming based on low-rank
  and cross-correlation techniques,'' \emph{IEEE Transactions on Signal
  Processing}, vol.~64, no.~15, pp. 3919--3932, 2016.

\bibitem{lrcc}
H.~{Ruan} and R.~C. {de Lamare}, ``Distributed robust beamforming based on
  low-rank and cross-correlation techniques: Design and analysis,'' \emph{IEEE
  Transactions on Signal Processing}, vol.~67, no.~24, pp. 6411--6423, 2019.

\bibitem{mbthp}
K.~{Zu}, R.~C. {de Lamare}, and M.~{Haardt}, ``Multi-branch tomlinson-harashima
  precoding design for mu-mimo systems: Theory and algorithms,'' \emph{IEEE
  Transactions on Communications}, vol.~62, no.~3, pp. 939--951, 2014.

\bibitem{rmbthp}
L.~{Zhang}, Y.~{Cai}, R.~C. {de Lamare}, and M.~{Zhao}, ``Robust multibranch
  tomlinson-harashima precoding design in amplify-and-forward mimo relay
  systems,'' \emph{IEEE Transactions on Communications}, vol.~62, no.~10, pp.
  3476--3490, 2014.

\bibitem{rsthp}
A.~R. {Flores}, R.~C. {De Lamare}, and B.~{Clerckx}, ``Tomlinson-harashima
  precoded rate-splitting with stream combiners for mu-mimo systems,''
  \emph{IEEE Transactions on Communications}, pp. 1--1, 2021.

\bibitem{bbprec}
L.~T.~N. {Landau} and R.~C. {de Lamare}, ``Branch-and-bound precoding for
  multiuser mimo systems with 1-bit quantization,'' \emph{IEEE Wireless
  Communications Letters}, vol.~6, no.~6, pp. 770--773, 2017.

\bibitem{baplnc}
J.~{Gu}, R.~C. {de Lamare}, and M.~{Huemer}, ``Buffer-aided physical-layer
  network coding with optimal linear code designs for cooperative networks,''
  \emph{IEEE Transactions on Communications}, vol.~66, no.~6, pp. 2560--2575,
  2018.

\bibitem{jpba}
Y.~{Jiang}, Y.~{Zou}, H.~{Guo}, T.~A. {Tsiftsis}, M.~R. {Bhatnagar}, R.~C. {de
  Lamare}, and Y.~D. {Yao}, ``Joint power and bandwidth allocation for
  energy-efficient heterogeneous cellular networks,'' \emph{IEEE Transactions
  on Communications}, vol.~67, no.~9, pp. 6168--6178, 2019.

\bibitem{zfsec}
X.~{Lu} and R.~C. d.~{Lamare}, ``Opportunistic relaying and jamming based on
  secrecy-rate maximization for multiuser buffer-aided relay systems,''
  \emph{IEEE Transactions on Vehicular Technology}, vol.~69, no.~12, pp.
  15\,269--15\,283, 2020.

\bibitem{rprec&pa}
V.~{M. T. Palhares}, A.~{Flores}, and R.~C. {De Lamare}, ``Robust mmse
  precoding and power allocation for cell-free massive mimo systems,''
  \emph{IEEE Transactions on Vehicular Technology}, pp. 1--1, 2021.

\bibitem{Nayebi2017}
E.~Nayebi, A.~Ashikhmin, T.~L. Marzetta, H.~Yang, and B.~D. Rao, ``Precoding
  and power optimization in cell-free massive {MIMO} systems,'' \emph{{IEEE}
  Transactions on Wireless Communications}, vol.~16, no.~7, pp. 4445--4459, jul
  2017.

\bibitem{itap&prec}
\BIBentryALTinterwordspacing
V.~M. Palhares, ``\BIBforeignlanguage{English}{Iterative ap selection, mmse
  precoding and power allocation in cell-free massive mimo systems},''
  \emph{\BIBforeignlanguage{English}{IET Communications}}, vol.~14, pp.
  3996--4006(10), December 2020. [Online]. Available:
  \url{https://digital-library.theiet.org/content/journals/10.1049/iet-com.2020.0627}
\BIBentrySTDinterwordspacing

\bibitem{Bjornson2020a}
E.~Bj\"{o}rnson and L.~Sanguinetti, ``Scalable cell-free massive {MIMO}
  systems,'' \emph{{IEEE} Transactions on Communications}, vol.~68, no.~7, pp.
  4247--4261, jul 2020.

\bibitem{Bjornson2020}
------, ``Making cell-free massive {MIMO} competitive with {MMSE} processing
  and centralized implementation,'' \emph{{IEEE} Transactions on Wireless
  Communications}, vol.~19, no.~1, pp. 77--90, jan 2020.

\bibitem{Interdonato2020}
G.~Interdonato, M.~Karlsson, E.~Bj\"{o}rnson, and E.~G. Larsson, ``Local
  partial zero-forcing precoding for cell-free massive {MIMO},'' \emph{{IEEE}
  Transactions on Wireless Communications}, vol.~19, no.~7, pp. 4758--4774, jul
  2020.

\bibitem{Buzzi2020}
S.~Buzzi, C.~D{\textquotesingle}Andrea, A.~Zappone, and
  C.~D{\textquotesingle}Elia, ``User-centric 5{G} cellular networks: Resource
  allocation and comparison with the cell-free massive {MIMO} approach,''
  \emph{{IEEE} Transactions on Wireless Communications}, vol.~19, no.~2, pp.
  1250--1264, feb 2020.

\bibitem{Li2003}
J.~Li, P.~Stoica, and Z.~Wang, ``On robust capon beamforming and diagonal
  loading,'' \emph{{IEEE} Transactions on Signal Processing}, vol.~51, no.~7,
  pp. 1702--1715, jul 2003.

\bibitem{Elnashar2006}
A.~Elnashar, S.~M. Elnoubi, and H.~A. El-Mikati, ``Further study on robust
  adaptive beamforming with optimum diagonal loading,'' \emph{{IEEE}
  Transactions on Antennas and Propagation}, vol.~54, no.~12, pp. 3647--3658,
  dec 2006.

\bibitem{Cai2015}
Y.~{Cai}, R.~C. {de Lamare}, L.~{Yang}, and M.~{Zhao}, ``Robust {MMSE}
  precoding based on switched relaying and side information for multiuser
  {MIMO} relay systems,'' \emph{{IEEE} Transactions on Vehicular Technology},
  vol.~64, no.~12, pp. 5677--5687, dec 2015.

\bibitem{Besson2005}
O.~Besson and F.~Vincent, ``Performance analysis of beamformers using
  generalized loading of the covariance matrix in the presence of random
  steering vector errors,'' \emph{{IEEE} Transactions on Signal Processing},
  vol.~53, no.~2, pp. 452--459, feb 2005.

\bibitem{Wang2013}
J.~Wang, M.~Bengtsson, B.~Ottersten, and D.~P. Palomar, ``Robust {MIMO}
  precoding for several classes of channel uncertainty,'' \emph{{IEEE}
  Transactions on Signal Processing}, vol.~61, no.~12, pp. 3056--3070, jun
  2013.

\bibitem{Tajer2011}
A.~Tajer, N.~Prasad, and X.~Wang, ``Robust linear precoder design for
  multi-cell downlink transmission,'' \emph{{IEEE} Transactions on Signal
  Processing}, vol.~59, no.~1, pp. 235--251, jan 2011.

\bibitem{Shen2012}
C.~{Shen}, T.~{Chang}, K.~{Wang}, Z.~{Qiu}, and C.~{Chi}, ``Distributed robust
  multicell coordinated beamforming with imperfect {CSI}: An {ADMM} approach,''
  \emph{{IEEE} Transactions on Signal Processing}, vol.~60, no.~6, pp.
  2988--3003, jun 2012.

\bibitem{Guo2005}
Y.~{Guo} and B.~C. {Levy}, ``Worst-case {MSE} precoder design for imperfectly
  known {MIMO} communications channels,'' \emph{{IEEE} Transactions on Signal
  Processing}, vol.~53, no.~8, pp. 2918--2930, aug 2005.

\bibitem{Wang2011}
J.~Wang and M.~Bengtsson, ``Joint optimization of the worst-case robust {MMSE}
  {MIMO} transceiver,'' \emph{{IEEE} Signal Processing Letters}, vol.~18,
  no.~5, pp. 295--298, may 2011.

\bibitem{Shen2014}
H.~Shen, J.~Wang, W.~Xu, Y.~Rong, and C.~Zhao, ``A worst-case robust {MMSE}
  transceiver design for nonregenerative {MIMO} relaying,'' \emph{{IEEE}
  Transactions on Wireless Communications}, vol.~13, no.~2, pp. 695--709, feb
  2014.

\bibitem{Ruan2019}
H.~Ruan and R.~C. de~Lamare, ``Distributed robust beamforming based on low-rank
  and cross-correlation techniques: Design and analysis,'' \emph{IEEE
  Transactions on Signal Processing}, vol.~67, no.~24, pp. 6411--6423, 2019.

\bibitem{Medra2018}
M.~Medra and T.~N. Davidson, ``Low-complexity robust {MISO} downlink precoder
  design with per-antenna power constraints,'' \emph{{IEEE} Transactions on
  Signal Processing}, vol.~66, no.~2, pp. 515--527, jan 2018.

\bibitem{Ketseoglou2019}
T.~Ketseoglou and E.~Ayanoglu, ``Zero-forcing per-group precoding for robust
  optimized downlink massive {MIMO} performance,'' \emph{{IEEE} Transactions on
  Communications}, vol.~67, no.~10, pp. 6816--6828, oct 2019.

\bibitem{Zhang2010}
B.~Zhang, Z.~He, K.~Niu, and L.~Zhang, ``Robust linear beamforming for {MIMO}
  relay broadcast channel with limited feedback,'' \emph{{IEEE} Signal
  Processing Letters}, vol.~17, no.~2, pp. 209--212, feb 2010.

\bibitem{Joham2005}
M.~{Joham}, W.~{Utschick}, and J.~A. {Nossek}, ``Linear transmit processing in
  {MIMO} communications systems,'' \emph{{IEEE} Transactions on Signal
  Processing}, vol.~53, no.~8, pp. 2700--2712, aug 2005.

\bibitem{StephenBoyd2019}
S.~Boyd and L.~Vandenberghe, \emph{Convex Optimization}.\hskip 1em plus 0.5em
  minus 0.4em\relax Cambridge University Press, 2004.

\bibitem{Tang2001}
A.~Tang, J.~Sun, and K.~Gong, ``Mobile propagation loss with a low base station
  antenna for {NLOS} street microcells in urban area,'' in \emph{{IEEE} {VTS}
  53rd Vehicular Technology Conference, Spring 2001. Proceedings (Cat.
  No.01CH37202)}.\hskip 1em plus 0.5em minus 0.4em\relax {IEEE}, may 2001.

\bibitem{deLamare2003}
R.~C. {de Lamare} and R.~{Sampaio-Neto}, ``Adaptive mber decision feedback
  multiuser receivers in frequency selective fading channels,'' \emph{IEEE
  Communications Letters}, vol.~7, no.~2, pp. 73--75, 2003.

\bibitem{itic}
R.~C. {De Lamare}, R.~{Sampaio-Neto}, and A.~{Hjorungnes}, ``Joint iterative
  interference cancellation and parameter estimation for cdma systems,''
  \emph{IEEE Communications Letters}, vol.~11, no.~12, pp. 916--918, 2007.

\bibitem{spa}
R.~C. {De Lamare} and R.~{Sampaio-Neto}, ``Minimum mean-squared error iterative
  successive parallel arbitrated decision feedback detectors for ds-cdma
  systems,'' \emph{IEEE Transactions on Communications}, vol.~56, no.~5, pp.
  778--789, 2008.

\bibitem{cai2009}
Y.~{Cai} and R.~C. {de Lamare}, ``Space-time adaptive mmse multiuser decision
  feedback detectors with multiple-feedback interference cancellation for cdma
  systems,'' \emph{IEEE Transactions on Vehicular Technology}, vol.~58, no.~8,
  pp. 4129--4140, 2009.

\bibitem{jiols}
R.~C. {de Lamare} and R.~{Sampaio-Neto}, ``Reduced-rank space-time adaptive
  interference suppression with joint iterative least squares algorithms for
  spread-spectrum systems,'' \emph{IEEE Transactions on Vehicular Technology},
  vol.~59, no.~3, pp. 1217--1228, 2010.

\bibitem{jiomimo}
Y.~{Cai} and R.~C. {de Lamare}, ``Space-time adaptive mmse multiuser decision
  feedback detectors with multiple-feedback interference cancellation for cdma
  systems,'' \emph{IEEE Transactions on Vehicular Technology}, vol.~58, no.~8,
  pp. 4129--4140, 2009.

\bibitem{mfdf}
P.~{Li} and R.~C. {De Lamare}, ``Adaptive decision-feedback detection with
  constellation constraints for mimo systems,'' \emph{IEEE Transactions on
  Vehicular Technology}, vol.~61, no.~2, pp. 853--859, 2012.

\bibitem{mbdf}
R.~C. {de Lamare}, ``Adaptive and iterative multi-branch mmse decision feedback
  detection algorithms for multi-antenna systems,'' \emph{IEEE Transactions on
  Wireless Communications}, vol.~12, no.~10, pp. 5294--5308, 2013.

\bibitem{did}
P.~{Li} and R.~C. {de Lamare}, ``Distributed iterative detection with reduced
  message passing for networked mimo cellular systems,'' \emph{IEEE
  Transactions on Vehicular Technology}, vol.~63, no.~6, pp. 2947--2954, 2014.

\bibitem{rrmser}
Y.~{Cai}, R.~C. {de Lamare}, B.~{Champagne}, B.~{Qin}, and M.~{Zhao},
  ``Adaptive reduced-rank receive processing based on minimum symbol-error-rate
  criterion for large-scale multiple-antenna systems,'' \emph{IEEE Transactions
  on Communications}, vol.~63, no.~11, pp. 4185--4201, 2015.

\bibitem{jidf}
R.~C. {de Lamare} and R.~{Sampaio-Neto}, ``Adaptive reduced-rank processing
  based on joint and iterative interpolation, decimation, and filtering,''
  \emph{IEEE Transactions on Signal Processing}, vol.~57, no.~7, pp.
  2503--2514, 2009.

\bibitem{bfidd}
A.~G.~D. {Uchoa}, C.~T. {Healy}, and R.~C. {de Lamare}, ``Iterative detection
  and decoding algorithms for mimo systems in block-fading channels using ldpc
  codes,'' \emph{IEEE Transactions on Vehicular Technology}, vol.~65, no.~4,
  pp. 2735--2741, 2016.

\bibitem{1bitidd}
Z.~{Shao}, R.~C. {de Lamare}, and L.~T.~N. {Landau}, ``Iterative detection and
  decoding for large-scale multiple-antenna systems with 1-bit adcs,''
  \emph{IEEE Wireless Communications Letters}, vol.~7, no.~3, pp. 476--479,
  2018.

\bibitem{detection_review}
R.~B. {Di Renna}, C.~{Bockelmann}, R.~C. {de Lamare}, and A.~{Dekorsy},
  ``Detection techniques for massive machine-type communications: Challenges
  and solutions,'' \emph{IEEE Access}, vol.~8, pp. 180\,928--180\,954, 2020.

\bibitem{aaidd}
R.~B. {Di Renna} and R.~C. {de Lamare}, ``Adaptive activity-aware iterative
  detection for massive machine-type communications,'' \emph{IEEE Wireless
  Communications Letters}, vol.~8, no.~6, pp. 1631--1634, 2019.

\bibitem{listmtc}
------, ``Iterative list detection and decoding for massive machine-type
  communications,'' \emph{IEEE Transactions on Communications}, vol.~68,
  no.~10, pp. 6276--6288, 2020.

\bibitem{dynmtc}
------, ``Dynamic message scheduling based on activity-aware residual belief
  propagation for asynchronous mmtc,'' \emph{IEEE Wireless Communications
  Letters}, pp. 1--1, 2021.

\bibitem{mwc}
F.~L. {Duarte} and R.~C. {de Lamare}, ``Cloud-driven multi-way multiple-antenna
  relay systems: Joint detection, best-user-link selection and analysis,''
  \emph{IEEE Transactions on Communications}, vol.~68, no.~6, pp. 3342--3354,
  2020.

\bibitem{dynovs}
Z.~{Shao}, L.~T.~N. {Landau}, and R.~C. {de Lamare}, ``Dynamic oversampling for
  1-bit adcs in large-scale multiple-antenna systems,'' \emph{IEEE Transactions
  on Communications}, pp. 1--1, 2021.

\bibitem{dopeg}
C.~T. {Healy} and R.~C. {de Lamare}, ``Decoder-optimised progressive edge
  growth algorithms for the design of ldpc codes with low error floors,''
  \emph{IEEE Communications Letters}, vol.~16, no.~6, pp. 889--892, 2012.

\bibitem{memd}
------, ``Design of ldpc codes based on multipath emd strategies for
  progressive edge growth,'' \emph{IEEE Transactions on Communications},
  vol.~64, no.~8, pp. 3208--3219, 2016.

\bibitem{vfap}
J.~{Liu} and R.~C. {de Lamare}, ``Low-latency reweighted belief propagation
  decoding for ldpc codes,'' \emph{IEEE Communications Letters}, vol.~16,
  no.~10, pp. 1660--1663, 2012.

\end{thebibliography}

\vfill


\end{document}